\documentclass{iau} 
\usepackage{graphicx}
\usepackage{natbib}

\newcommand{\solphys}{{\it Solar Phys.}}
\newcommand{\apj}{{\it Astrophys. J.}}
\newcommand{\pasp}{{\it Pub. Astron. Soc. Pac.}}
\newcommand{\apjl}{{\it Astrophys. J. Lett.}}
\newcommand{\pasj}{{\it Pub. Astron. Soc. Japan}}
\newcommand{\aap}{{\it Astron. Astrophys.}}
\newcommand{\grl}{{\it Geophys. Res. Lett.}}
\newcommand{\apjs}{{\it Astrophys. J. Supp. Series}}

\title[Stellar Coronal Dimming] 
{Coronal Dimming as a Proxy for Stellar Coronal Mass Ejections}

\author[M. Jin et al.]   
{M. Jin$^{1, 2}$, M. C. M. Cheung$^{2, 3}$, M. L. DeRosa$^2$, N. V. Nitta$^2$, C. J. Schrijver$^2$, K. France$^4$, A. Kowalski$^4$, J. P. Mason$^4$, R. Osten$^5$}

\affiliation{$^1$SETI Institute, \\ 
189 N Bernardo Ave suite 200, Mountain View, CA 94043, USA \\ 
email: {\tt jinmeng@lmsal.com} \\[\affilskip]
$^2$Lockheed Martin Solar and Astrophysics Lab (LMSAL), \\ 
3251 Hanover St., Bldg. 252, Palo Alto, CA 94304, USA \\[\affilskip]
$^3$Hansen Experimental Physics Laboratory, Stanford University, \\
452 Lomita Mall, Stanford, CA 94305, USA \\[\affilskip]
$^4$Laboratory for Atmospheric and Space Physics, University of Colorado at Boulder, \\
1234 Innovation Dr, Boulder, CO 80303, USA \\[\affilskip]
$^5$Space Telescope Science Institute, \\
3700 San Martin Dr, Baltimore, MD 21218, USA}
\pubyear{2019}
\volume{354}  
\jname{Solar and Stellar Magnetic Fields: Origins and Manifestations}
\editors{Kosovichev, A., Strassmeier, K. G. \& Jardine, M., eds.}
\begin{document}

\maketitle

\begin{abstract}
Solar coronal dimmings have been observed extensively in the past two decades and are believed to have close association with coronal mass ejections (CMEs). Recent study found that coronal dimming is the only signature that could differentiate powerful flares that have CMEs from those that do not. Therefore, dimming might be one of the best candidates to observe the stellar CMEs on distant Sun-like stars. In this study, we investigate the possibility of using coronal dimming as a proxy to diagnose stellar CMEs. By simulating a realistic solar CME event and corresponding coronal dimming using a global magnetohydrodynamics model (AWSoM: Alfv\'{e}n-wave Solar Model), we first demonstrate the capability of the model to reproduce solar observations. We then extend the model for simulating stellar CMEs by modifying the input magnetic flux density as well as the initial magnetic energy of the CME flux rope. Our result suggests that with improved instrument sensitivity, it is possible to detect the coronal dimming signals induced by the stellar CMEs.

\end{abstract}

\firstsection 
\section{Introduction}
``Coronal dimming" refers as the reduction in intensity on or near the solar disk across a large area during solar eruptive events. It was first observed in white light corona and described as a ``depletion" \citep{hansen74} and later was found in solar X-ray observations as ``transient coronal holes" \citep{rust76}. The studies about coronal dimming dramatically increased with Solar and Heliospheric Observatory (SOHO)/Extreme-ultraviolet Imaging (EIT) observations \citep{dela95}, with which the coronal dimming was first observed in multiple EUV channels with different emission temperatures \citep{thompson98}. The coronal dimming was also found usually associated with coronal EUV waves (also called ``EIT waves", \citealt{thompson99}). Recently, with EUV observations of unprecedented high temporal ($\sim$12 s) and spatial resolution ($\sim$0.6 arcsec) in seven channels from Solar Dynamics Observatory (SDO; \citealt{pesnell12})/Atmospheric Imaging Assembly (AIA; \citealt{lemen12}) as well as the high spectral resolution data from SDO/Extreme Ultraviolet Variability Experiment (EVE; \citealt{woods12}), it provides us an unique opportunity for in-depth studies about coronal dimming. 

Two decades of solar observations suggest that all coronal dimmings are associated with coronal mass ejections (CMEs; e.g., \citealt{sterling97, reinard08}). Furthermore, since observations show simultaneous and co-spatial dimming in multiple coronal lines (e.g., \citealt{zarro99, sterling00}) and the spectroscopic observations show that the dimming region has up-flowing expanding plasma (e.g., \citealt{harra01, harra07, imada07, jin09, attrill10, tian12}), it is widely accepted that the coronal dimming is due to the plasma evacuation during the CME and the dimming area is believed to be the footpoints of the erupting flux rope. Recent magnetohydrodynamics (MHD) modeling results of coronal dimming (e.g., \citealt{cohen09,downs12}) also suggest that the dimming is mainly caused by the CME-induced plasma evacuation, and the spatial location is well correlated with the footpoints of the erupting magnetic flux system \citep{downs15}. Although, there are other known mechanisms that could cause coronal dimming in observations \citep{mason14}.

Due to its close association with CMEs, coronal dimming encodes important information about CME’s mass, speed, energy etc. (e.g., \citealt{hudson96, sterling97, harrison03, zhukov04, asch09, cheng16, krista17, dissauer18a, dissauer18b}), therefore provide critical estimations for space weather forecast. For example, \citet{krista13} found correlations between the magnitudes of dimmings/flares and the CME mass by studying variation between recurring eruptions and dimmings. Using SDO/EVE observations, \citet{mason16} found that the CME velocity/mass can be related to coronal dimming properties (e.g., dimming depth and dimming slope), which could be used to estimate CME mass and speed in the space weather forecast operations. On the other hand, by exploring the characteristics of 42 X-class solar flares, \citet{harra16} found that coronal dimming is the only signature that could differentiate powerful flares that have CMEs from those that do not. Therefore, dimming might be one of the best candidates to observe the stellar CMEs on distant Sun-like stars. 

In this study, we first model the coronal dimming associated with a realistic CME event on 2011 February 15 \citep{schrijver11, jin16} to demonstrate the capability of the model to reproduce solar observations. We then extend the model for simulating stellar CMEs by changing the input magnetic flux density as well as the initial energy of the CME flux rope.  In \S 2 we briefly introduce the global MHD model used in this study, followed by the results and discussion in \S 3.

\section{Global Corona \& CME Models}
In this study, the model for reconstructing the global corona environment is the Alfv\'{e}n Wave Solar Model \citep{bart14} within the Space Weather Modeling Framework (SWMF; \citealt{toth12}). AWSoM is a data-driven global MHD model with inner boundary specified by observed magnetic maps and simulation domain extending from the upper chromosphere to the corona and heliosphere. Physical processes included in the model are multi-species thermodynamics, electron heat conduction (both collisional and collisionless formulations), optically thin radiative cooling, and Alfv\'{e}n-wave turbulence that accelerates and heats the solar wind. The Alfv\'{e}n-wave description includes non-Wentzel-Kramers-Brillouin (WKB) reflection and physics-based apportioning of turbulent dissipative heating to both electrons and protons. AWSoM has demonstrated the capability to reproduce high-fidelity solar corona conditions \citep{sokolov13, bart14, oran13, oran15, jin16, jin17a}.

Based on the steady-state global corona solution, we initiate the CME by using an analytical Gibson-Low (GL) flux rope model \citep{gibson98}. This flux rope model has been successfully used in numerous modeling studies of CMEs (e.g., \citealt{chip04a, chip04b, lugaz05, chip14, jin16, jin17a}). \citet{jin17b} developed a module (EEGGL) to calculate the GL flux rope parameters based on near-Sun observations so that this first-principles-based MHD model could be utilized as a forecasting tool. Analytical profiles of the GL flux rope are obtained by finding a solution to the magnetohydrostatic equation $(\nabla\times{\bf B})\times{\bf B}-\nabla p-\rho {\bf g}=0$ and the solenoidal condition $\nabla\cdot{\bf B}=0$. After inserting the flux rope into the steady-state solar corona solution: i.e. $\rho=\rho_{0}+\rho_{\rm GL}$, ${\bf B = B_{0}+B_{\rm GL}}$, $p=p_{0}+p_{\rm GL}$, the combined background-flux-rope system is in a state of force imbalance and thus erupts immediately when the numerical model is advanced forward in time. The simulation setup in this study is similar to that in our previous work \citep{jin16}. 

\section{Results \& Discussion}
To demonstrate the capability of the model for reproducing the solar coronal dimming, we synthesize the EUV emissions of 6 AIA wavebands for two hours after the CME onset in the 2011 February 15 event \citep{schrijver11, jin16}. In addition, we calculate the Emission Measure (EM) in 4 temperature bins ($5.75<lgT<6.55$). In Figure 1, we show the core dimming (near the source region) evolution in the simulation. The left panel in Figure 1 shows a base-difference image (by substracting the pre-event intensity) in AIA 211 \AA~at $t$ = 1 hour. The black box shows the sub-region where the EUV intensities and EMs are derived. In the middle panel of Figure 1, the EUV intensity evolution (represented by relative changes in percentage comparing with the pre-event values) in 2 hours for all 6 AIA wavebands is shown. We can see that the intensities in all 6 wavebands drop, which suggests the plasma depletion included by the CME. This is also evident in the EM calculation (right panel of Figure 1). By fitting the intensity curves, we can estimate the dimming recovery time is about $\sim$9 to 16 hours.
\begin{figure}[htb]
\begin{center}
 \includegraphics[width=5.4in]{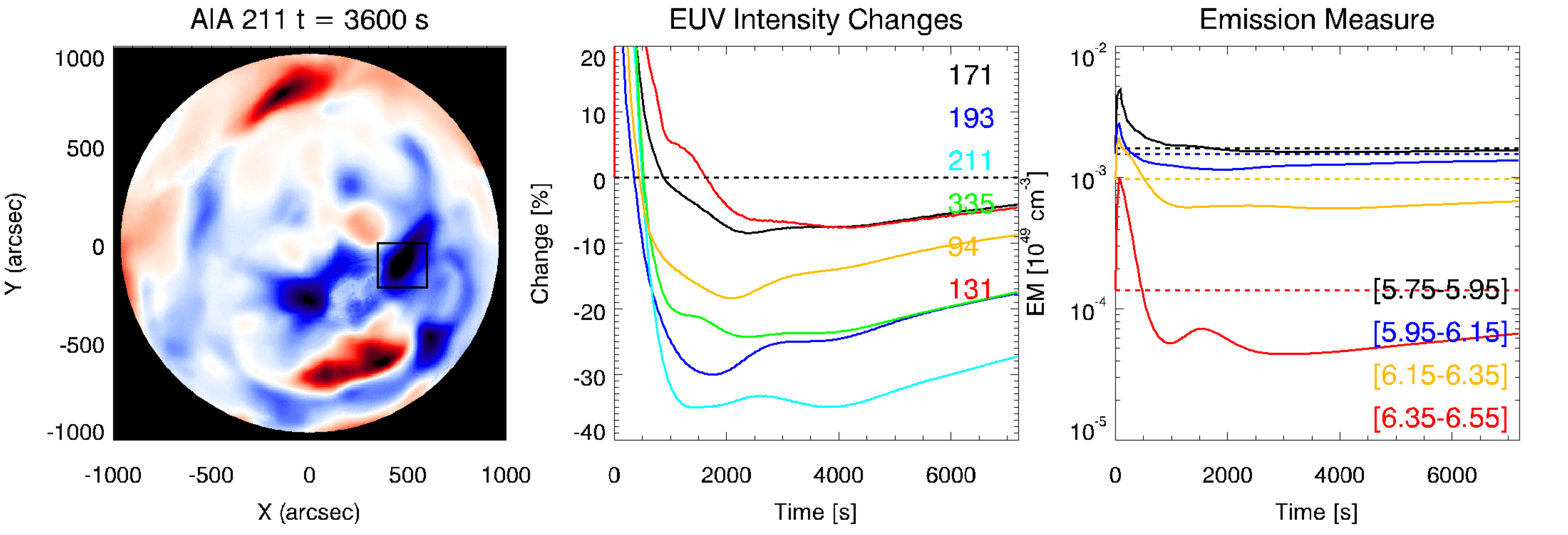} 
 \caption{Coronal dimming evolution in the simulation of 2011 February 15 event. Left panel: synthesized AIA 211 $\AA$ base difference image at t = 1 hour. The black box shows the sub-region where the EUV intensity and Emission Measure (EM) are derived. Middle panel: the EUV intensity changes in 6 synthesized AIA channels. Right panel: EM evolution in the simulation. The dashed lines show the pre-event EM value.}
 \label{fig1}
\end{center}
\end{figure}

With higher mean surface magnetic flux density $\langle|fB|\rangle$ on the M dwarf or young Sun-like stars, their coronas are believed to be hotter than the solar case. By varying the $\langle|fB|\rangle$ in the model by 5, 10, 20, 30 times, we demonstrate this effect quantitatively in Figure 2. With the increasing magnetic field strength, the peak EM temperature shifts from $\sim$1 MK for the solar case to $\sim$5 MK for the stellar case with magnetic field 30 times stronger. Note that the peak EM temperature for the solar case is consistent with the dominant dimming lines in the solar observation (i.e., 171 \AA~and 193 \AA, with emission temperatures around 1 MK). Therefore, even without simulating the CMEs, the steady-state corona solution is useful to estimate the dominant coronal dimming lines in the stellar cases with different magnetic flux densities.
\begin{figure}[htb]
\begin{center}
 \includegraphics[width=5.0in]{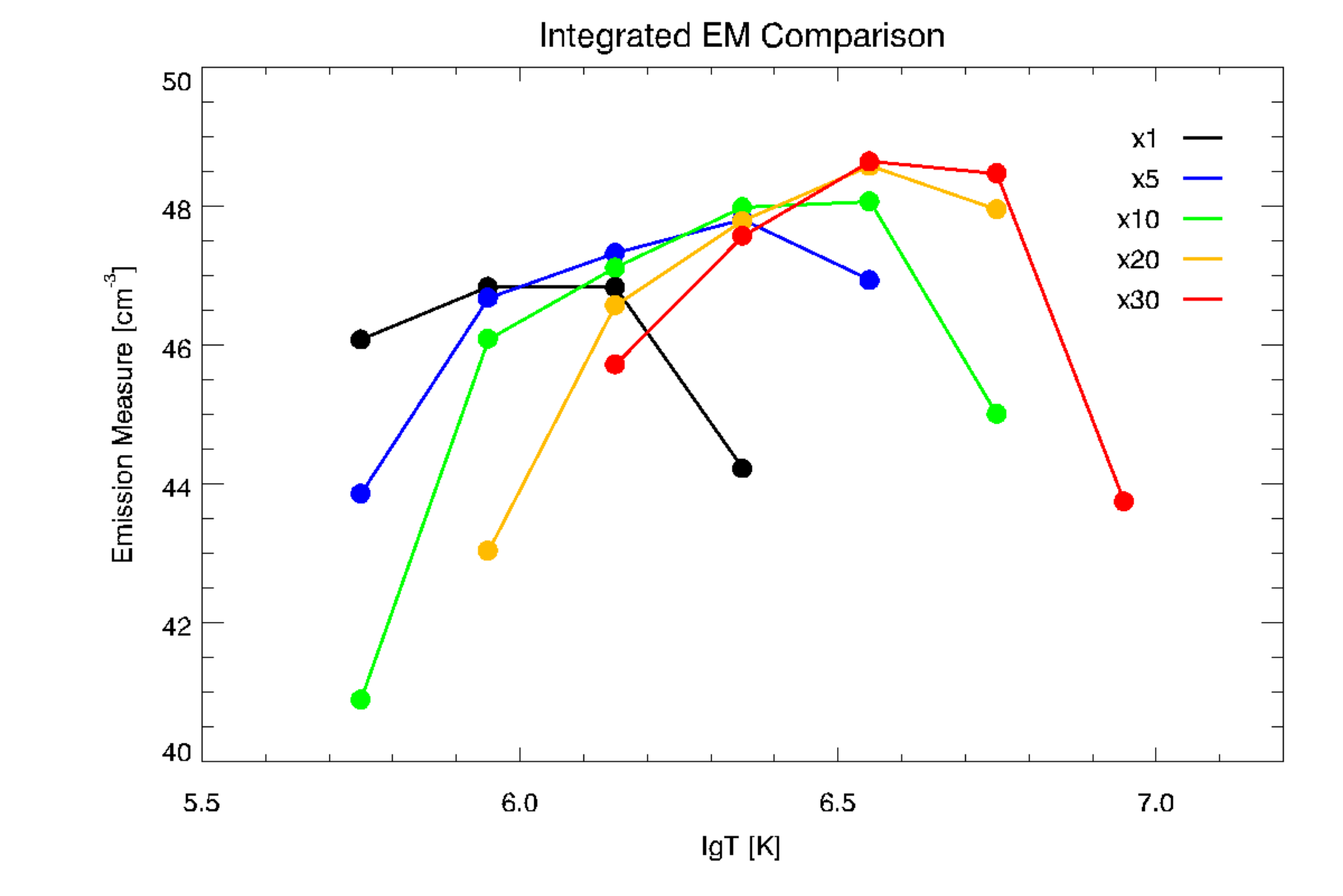} 
 \caption{Integrated EM for different input magnetic flux densities. ``x1" represents the solar case, and ``x30" means the magnetic flux density is 30 times the solar case.}
   \label{fig2}
\end{center}
\end{figure}

We then initiate CMEs with varying initial energies and into different steady-state coronal conditions. The simulation is switched to time-accurate mode to capture the CME eruption, and the MHD equations are solved in conservative form to guarantee the energy conservation during the eruption process. Here, we show two representative cases: one confined eruption and one explosive eruption. The confined eruption is due to the strong overlaying global coronal magnetic field that prevents the CME to escape \citep{alvarado18}. In this case, since there is no plasma depletion involved, the EUV lines show no dimming features. Instead, due to the redistribution of the magnetic energy in the corona, it leads to a second peak in some EUV intensity profiles as shown in Figure 3. The second peak is more evident in the lines with higher emission temperature (e.g., 335 \AA~and 284 \AA), which suggests the coronal plasma is dominant in these temperatures. This phenomenon is also frequently observed on the Sun and referred as EUV late-phase (e.g., \citealt{woods11}).
\begin{figure}[htb]
\begin{center}
 \includegraphics[width=4.8in]{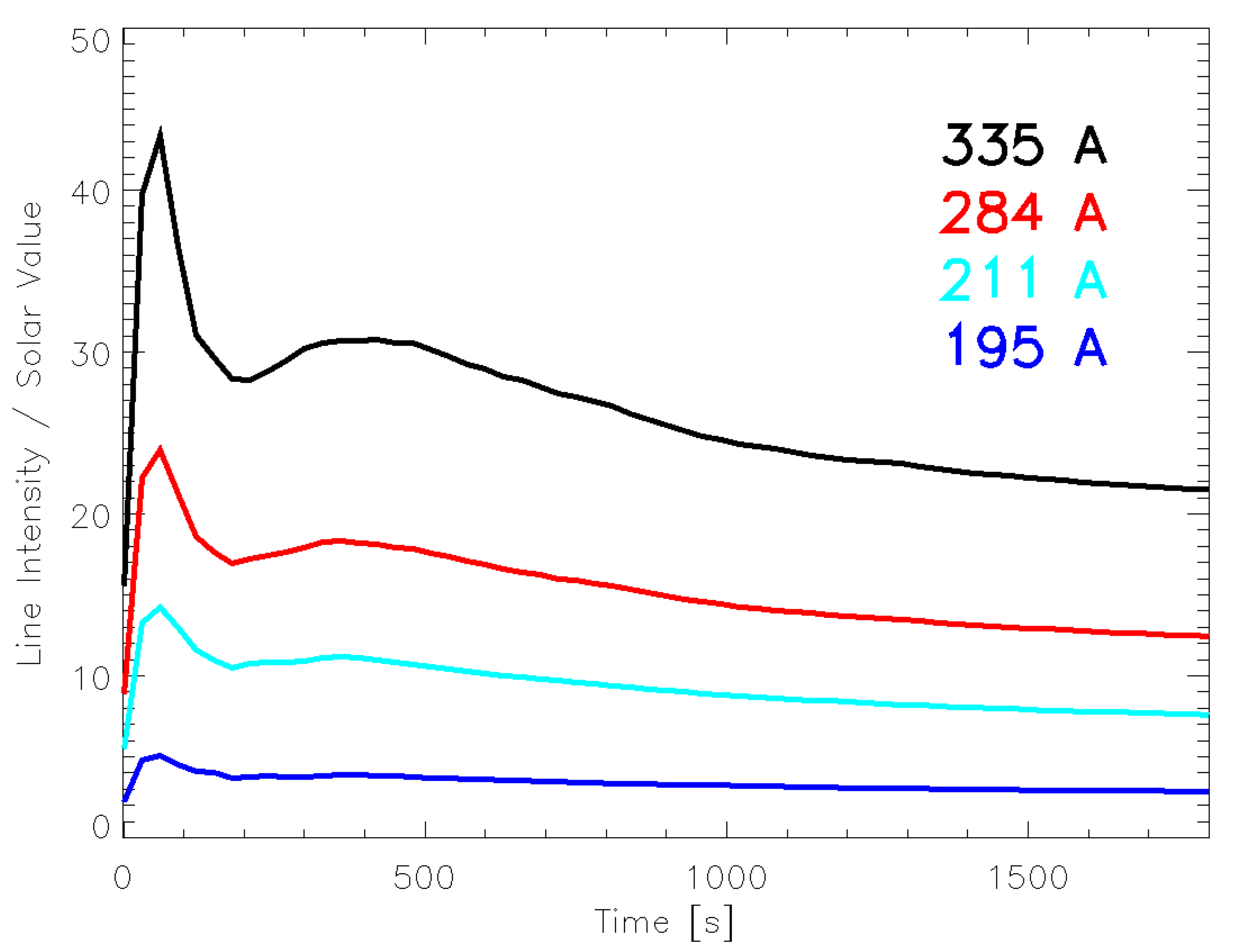} 
 \caption{The EUV intensity evolution of 4 different wavelengths for a confined eruption case. The intensities are scaled with the steady-state solar intensities.}
   \label{fig3}
\end{center}
\end{figure}

The second case is an explosive eruption in which the coronal mass is released into the interstellar space and leads to EUV coronal dimmings. In this case, the input $\langle|fB|\rangle$ is 5 times higher than the solar case. And the initial CME flux rope energy is about 10$^{33}$ ergs. The resulting CME speed is $\sim$3000 km s$^{-1}$, which is only slightly higher than the fast solar CMEs due to the stronger coronal field confinement. We simulate the coronal dynamical evolution for 4 hours after the CME eruption and synthesize the EUV line emissions. To further investigate the detectability of the dimming signals, we apply the instrument performance estimates from the Extreme-ultraviolet Stellar Characterization for Atmospheric Physics and Evolution (ESCAPE) mission concept, which provides extreme- and far-ultraviolet spectroscopy (70 - 1800 \AA) to characterize the high energy radiation environment in the habitable zones around nearby stars \citep{france19}. The EUV detector of the ESCAPE mission will be $\sim$100 times more sensitive than the previous EUVE mission \citep{craig97}. Figure 4 shows the simulated Fe XV 284 \AA~and Fe XVI 335 \AA~line emissions by assuming the star at 6 pc and using 30 minutes exposure time. The ISM absorption effect has been taken into account assuming $N$(H I) = $10^{18}$ cm$^{-2}$ \citep{france18}. This preliminary result suggests that with better instrumentation, the stellar coronal dimmings associated with CMEs can be observed. 

\vspace{4mm}We summarize the main results as follows:
\begin{itemize}
	\item The coronal dimmings encode important information about CME energetics, CME-driven shock properties, and magnetic configuration of erupting flux ropes.
	
	\item With higher magnetic flux density, the stellar dimming occurs in the higher temperature range than the solar case. With better instrumentation, the coronal dimming could be detected from the distant stars.
	
	\item Our results show a proof-of-concept that the MHD model can be used for quantitative studies of CME-dimming relationships and applied to stellar cases. To get useful CME information from the future stellar observations, more detailed modeling studies are needed.
\end{itemize}
\begin{figure}[htb]
\begin{center}
 \includegraphics[width=4.5in]{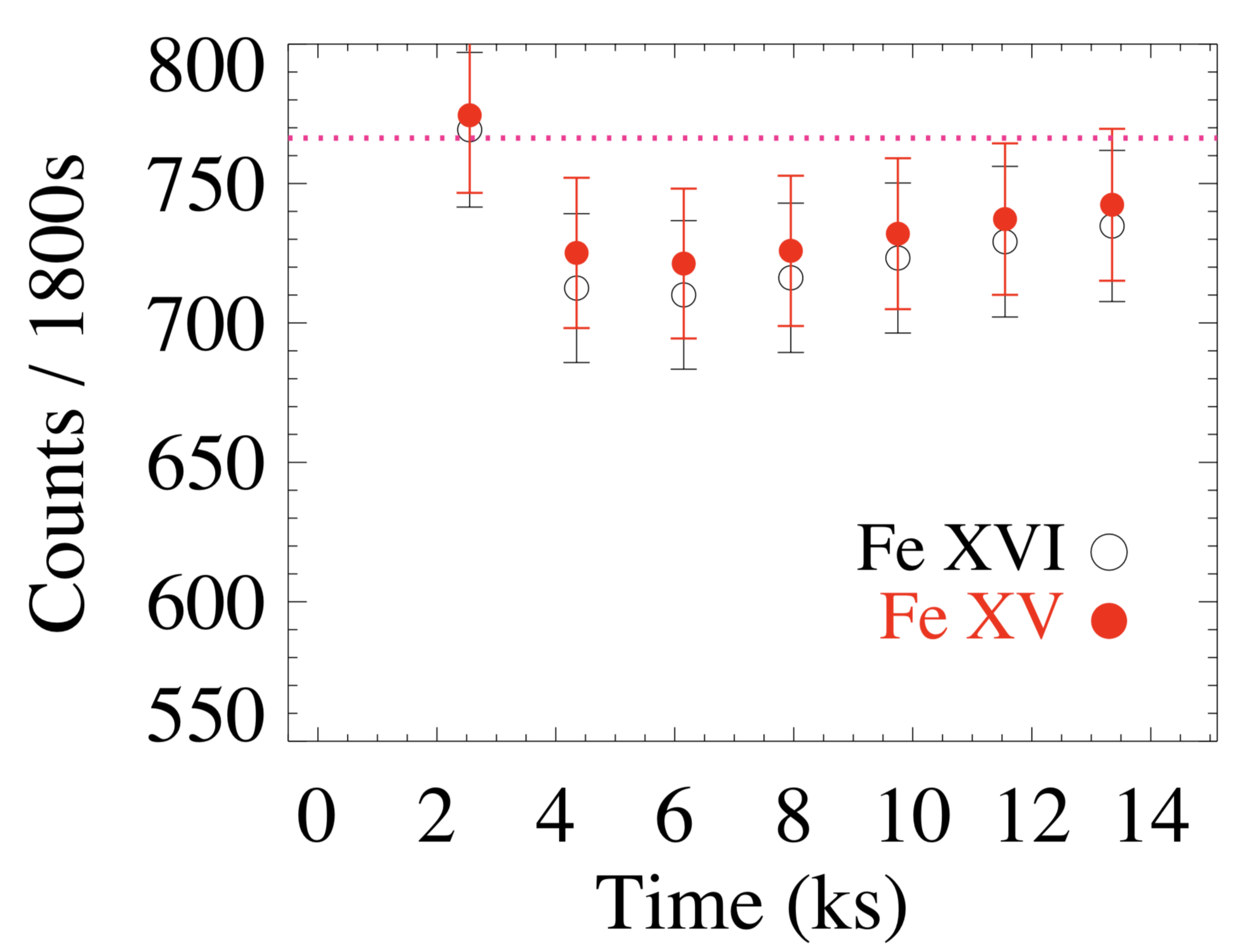} 
 \caption{Synthesized Fe XV 284 \AA~and Fe XVI 335 \AA~line intensity evolution after CME onset from ESCAPE instrument performance estimates. The detector counts are based on 30 minutes exposure time.}
   \label{fig4}
\end{center}
\end{figure}

\begin{discussion}
\discuss{Christoffer Karoff}{How would this look if you looked in the optical and not the UV?}

\discuss{Meng Jin}{Because the mass loss of the CME is mainly from the corona, the coronal dimming is not seen in the optical bands that dominate by photospheric emissions.}
\end{discussion}

\begin{acknowledgements}
M.Jin was supported by NASA's SDO/AIA contract (NNG04EA00C) to LMSAL. We are thankful for the use of the NASA Supercomputer Pleiades at Ames and its helpful staff for making it possible to perform the simulations presented in this study. SDO is the first mission of NASA’s Living With a Star (LWS) Program.
\end{acknowledgements}

\end{document}